\documentclass{IEEEtran}

\usepackage{breakurl}
\usepackage{multirow}
\usepackage{adjustbox}
\usepackage{enumitem}
\usepackage{graphicx}
\usepackage{colortbl}
\usepackage{amsmath}
\usepackage[font=sc]{caption}
\usepackage{url}

\usepackage{subcaption}
\usepackage{breqn}
\usepackage{enumitem}
\usepackage[ruled]{algorithm2e}
\usepackage{threeparttable}
\usepackage{lipsum}
\usepackage[mathlines,switch]{lineno}
\usepackage{booktabs}
\usepackage{pifont}
\usepackage{soul}
\usepackage{bm}
\usepackage{hyperref}
\usepackage{acronym}
\def\BibTeX{{\rm B\kern-.05em{\sc i\kern-.025em b}\kern-.08em
    T\kern-.1667em\lower.7ex\hbox{E}\kern-.125emX}}
\begin{document}
\acrodef{BD-BR}{Bjontegaard Delta - Bitrate}
\acrodef{PVS}{Processed Video Sequence}
\acrodef{QP}{quantization parameter}

\title{Codec Compression Efficiency Evaluation of MPEG-5 part 2 (LCEVC) using Objective and Subjective Quality Assessment}
\author{ Nabajeet~Barman, Steven Schmidt, Saman Zadtootaghaj, and Maria G.~Martini
\thanks{Nabajeet Barman and Maria Martini are with Wireless Multimedia \& Networking Research Group, Kingston University London, UK. Steven Schmidt and Saman Zadtootaghaj are with Quality and Usability Lab, TU Berlin, Germany. \newline
Email: Nabajeet.Barman@kingston.ac.uk, m.martini@kingston.ac.uk, steven.schmidt@tu-berlin.de,Saman.Zadtootaghaj@qu.tu-berlin.de}}

\markboth{Preprint of a manuscript currently under review}
{Shell \MakeLowercase{\textit{et al.}}: Bare Demo of IEEEtran.cls for IEEE Journals}

\maketitle

\begin{abstract}
With the increasing advancements in video compression efficiency achieved by newer codecs such as HEVC, AV1, and VVC, and intelligent encoding strategies, as well as improved bandwidth availability,there has been a proliferation and acceptance of newer services such as Netflix, Twitch, etc. However, such higher compression efficiencies are achieved at the cost of higher complexity and encoding delay, while many applications are delay sensitive. Hence, there is a requirement for faster, more efficient codecs to achieve higher encoding efficiency without significant trade-off in terms of both complexity and speed. We present in this work an evaluation of the latest MPEG-5 Part 2 Low Complexity Enhancement Video Coding (LCEVC) for live gaming video streaming applications. The results are presented in terms of bitrate savings using both subjective and objective quality measures as well as a comparison of the encoding speeds. Our results indicate that, for the encoding settings used in this work, LCEVC outperforms both x264 and x265 codecs in terms of bitrate savings using VMAF by approximately 42\% and 38\%. Using subjective results, it is found that LCEVC outperforms the respective base codecs, especially for low bitrates. This effect is more evident for x264 than for x265, i.e., for the latter the absolute improvement of quality scores is smaller. The objective and subjective results as well as sample video sequences are made available as part of an open dataset, LCEVC-LiveGaming\footnote{\url{https://github.com/NabajeetBarman/LCEVC-LiveGaming}}.    
\end{abstract}


\section{Introduction}
In recent years both passive streaming applications such as Twitch \cite{Barman2019PhDThesis} and cloud gaming applications such as PS Now, Xbox Cloud Gaming, and Google Stadia 
have gained 
attention and popularity. In order to provide a reliable and pleasant experience, it is imperative for the service providers to ensure a reasonable end user Quality of Experience (QoE). However, low and often fluctuating bandwidth availabilities necessitate 
effective video compression so that the quality is gracefully degraded rather than being totally disrupted.

Over the years, many advancements 
led to
the development of more efficient, but complex codecs, such as AV1 by Alliance for Open Media and H.266/VVC as a successor of HEVC. These newer codecs achieve better compression ratios at the cost of high complexity and hence increased encoding times \cite{Ronca2019EncoderComplexity}. Live streaming is a complex process and is highly delay intolerant. Hence, often the codecs need to be run at 
high speed, often bypassing many of the 
more complex encoding steps, thus 
reducing
the actual performance gains 
compared to 
what is
reported for on-demand streaming applications.

More recently, in Oct 2020, MPEG finalized the MPEG-5 Part 2 Low Complexity Enhancement Video Coding (LCEVC) standard, publsihed in Nov 2021 as a new video coding standard as ISO/IEC 23094-2 \cite{LCEVC-MPEG-ISO}, which is shown to reduce the overall encoding complexity with negligible impact on the coding efficiency and, in several cases, even improvement in the coding efficiency \cite{LCEVC}. Additionally, it provides backwards compatibility with existing ecosystem of devices.

We present in this work an evaluation of the LCEVC codec for live gaming video streaming applications. The results are presented in terms of \ac{BD-BR}
\cite{MHV_BD_BR,BD_BR_original}
savings and subjective scores. 
The major contributions of this work are:
\begin{enumerate}
    \item Evaluation of the new MPEG-5 Low Complexity Enhancement Video Coding (LCEVC) standard on gaming content;
    \item Comparison of compression efficiency of LCEVC with existing practical implementations of the video codec standards H.264/AVC and H.265/HEVC in terms of bitrate savings using objective quality metrics (PSNR and VMAF);
    \item Comparison of the codecs using subjective video quality scores;
    \item An in-depth analysis on the performance of the LCEVC codec using per-frame quality scores;
    \item An open-source dataset to support the  reproducibility of the results presented in this paper as well as  additional future  research.
\end{enumerate}

The rest of the paper is organized as follows. Section~\ref{sec:Back} presents some information about the LCEVC codec and some related literature work. In Section~\ref{sec:DandEM} we present the dataset used and the evaluation methodology. Section~\ref{sec:Results} presents the results and we conclude the paper in Section~\ref{sec:Conc} with a discussion of future work.

\section{Background and Related Work} \label{sec:Back}

\subsection{Gaming Video Streaming}
 Gaming has been a prevalent form of entertainment for many years. Recently, there has been an increasing popularity of both passive and active gaming applications. Passive live gaming video streaming applications such as Twitch and Facebook Gaming 
 stream the gameplay of players for "passive" viewing by others. However, due to their chat function and other means of involving the community, those services are becoming more interactive as well. On the other hand, there has also been an increasing effort to establish highly interactive cloud gaming services which run a video game on a cloud server 
 and stream the generated scene as a video to the user, who in turn can control the game remotely. The interactive nature of these services make them very sensible to latency. We focus on in this work only on the video streaming part of such applications. For a more detailed discussion on these two gaming streaming application types, we refer the reader to \cite{Barman2019PhDThesis}.
 
\subsection{Video Codecs}
\begin{figure*}[t!]
\begin{center}
\includegraphics[ width=1.0\linewidth]{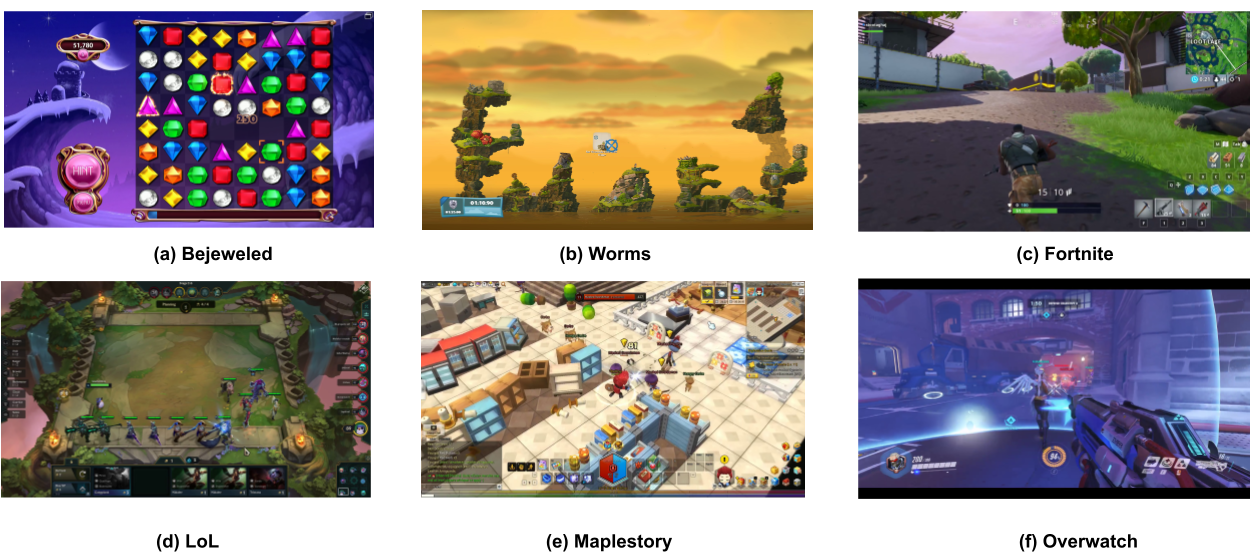}
\end{center} 
\caption{Snapshots of sample six games out of the total fourteen considered in this work.}
\label{fig:screenshots}
\end{figure*}

With a first version finalised in 2003,
H.264/AVC is one of the most widely used codecs worldwide, with approximately still 70\% of the videos being encoded using this standard. Its successor, H.265/HEVC, was standardized in 2013 with the aim to provide approximately 50\% higher compression efficiency vs. H.264, but since then has seen less adoption due to unclear royalty structure 
and licence fees. 
More recently standardized, LCEVC 
can use
any existing codec (e.g., H.264, H.265, AV1). It  encodes the videos at a lower resolution and 
uses novel approaches to compress the residuals, 
which are then used to correct the artefacts produced during the re-scaling of the encoded lower resolution video to native resolution. The enhancement achieved by using the residuals improves the details and sharpness of the final decoded video sequence. Increased speed and lower complexity are achieved due to fewer pixels for the base codec to encode and highly efficient and fast encoding of the residuals.

\subsection{Related Work}

Over the past 
years, codec comparison has been a very active field of research. While most of the works evaluate the compression efficiency achieved by the compression standard 
in general 
\cite{Topiwala2019CodecComparisonEVC} 
\cite{Laude2018CodecComparison}, others,
e.g.,
\cite{Netflix2016LargeScaleCodecComparison}, \cite{DarkhanAV1}, \cite{Zabrovskiy2018CodecComparison} and \cite{GamingHDRVideoSET}, evaluate the codec compression efficiency as achieved by the practical implementations of the video compression standards.

So far, few works have evaluated the compression efficiency of LCEVC for both on-Demand and live streaming applications \cite{LCEVC-MPEG,OTTVerse}. More recently, Jan Ozer \cite{JanOzerLiveUseCase} evaluated LCEVC's performance for live eSports and gaming applications. However, the work was limited to only one codec (x264). Towards this end, we present in this work the first 
independent academic evaluation of the newly proposed LCEVC codec for live gaming video streaming applications using an open-source dataset. In addition to \ac{BD-BR} results, we present an in-depth analysis on the performance of the codec using per-frame quality scores and frame buffer size. 

\section{Dataset and Evaluation Methodology} \label{sec:DandEM}

\subsection{Dataset}

In this work, we use 14 reference gaming video sequences from the publicly available Cloud Gaming Video DataSet (CGVDS) \cite{SamanCGVDS}. The reference video sequences are each of 30 seconds duration, 1920x1080 resolution, and 60 fps. Figure~\ref{fig:screenshots} presents snapshots of six sample games used in this work.

\begin{figure}[htb!]
\begin{center}
\includegraphics[ width=1.0\linewidth]{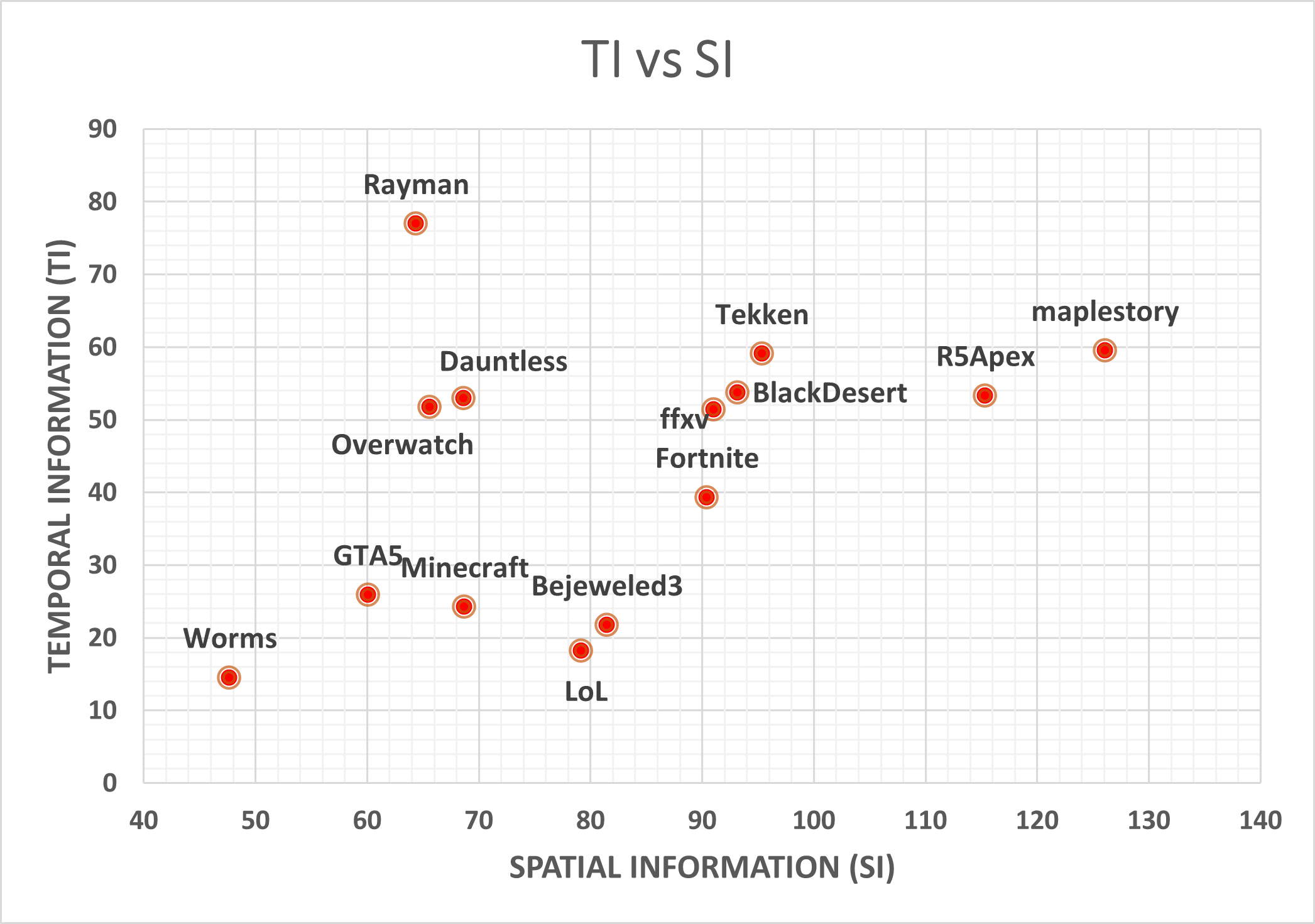}
\end{center} 
\caption{TI vs. SI plot of the fourteen games considered in this work.}
\label{fig:SITI}
\end{figure}

A plot of Spatial Information (SI) and Temporal Information (TI) values, as defined in ITU-T Rec P.910 \cite{itu1999subjective}, 
providing information on the
encoding complexity, is presented in Figure~\ref{fig:SITI}. The SI-TI values were calculated using the open-source tool available on Github \cite{SITIGithub_barman}. Based on the figure, it is clear that the selected games are from various genres and are of different complexity, and hence the results are generalizable to different content types.

\begin{figure*}[t!]
\begin{center}
\includegraphics[width=1.0\linewidth]{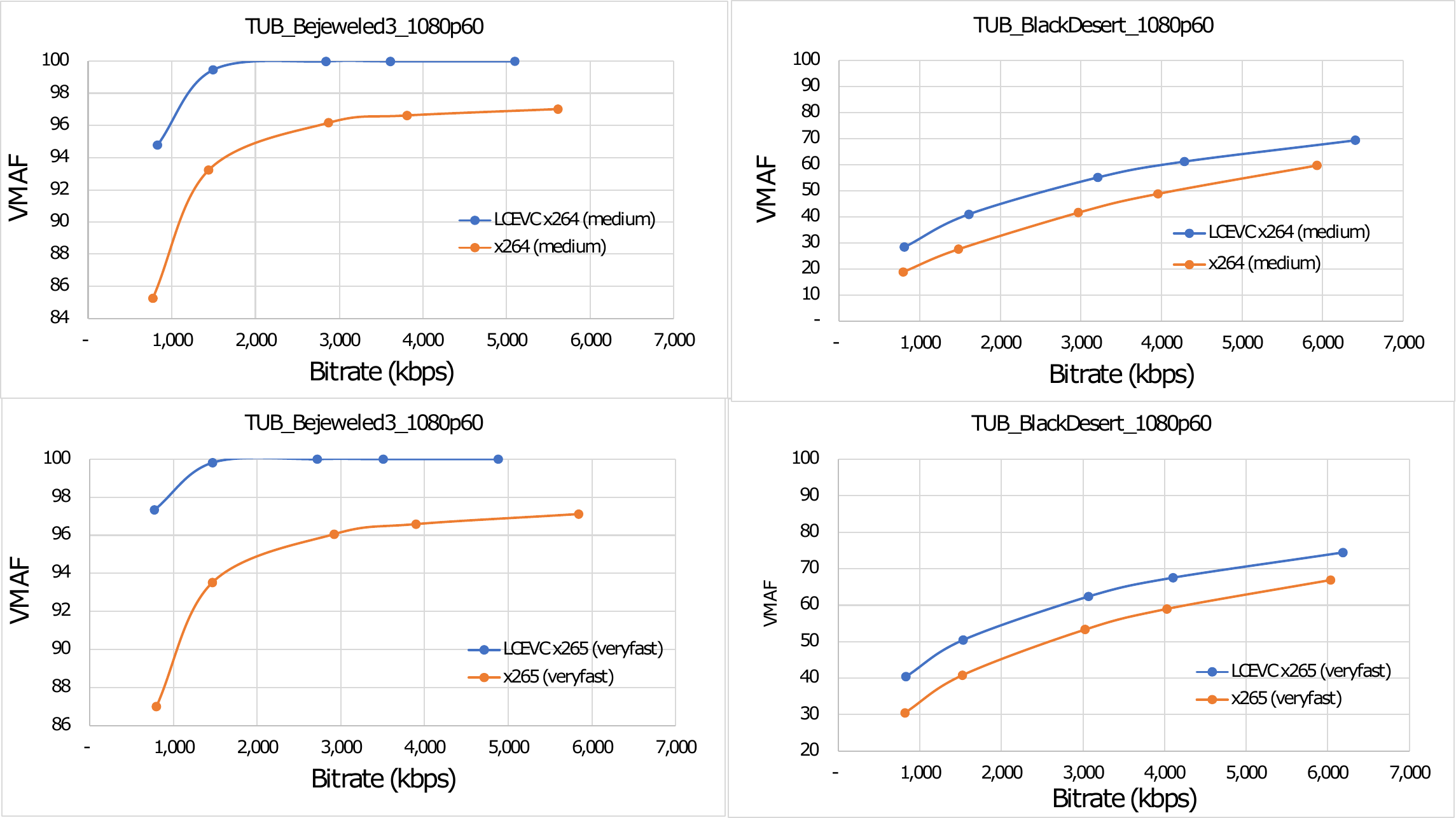}
\end{center} 
\caption{Quality (VMAF) vs. Bitrate (kbps) for two sample games considering both base codecs, x264 (frst row) and x265 (second row) and their respective LCEVC implementation (LCEVC-x264 and LCEVC-x265).}
\label{fig:QualityBRcurves}
\end{figure*}


\subsection{Evaluation Methodology}
In this work, we limit our analysis to single resolution-bitrate encoding settings. 
For encoding the reference videos, we used the SDK releases made available by V-Nova which include FFmpeg with LCEVC support for x264 and x265 as base codecs, referred to as LCEVC-x264 and LCEVC-x265, respectively. The reference videos were then encoded at the native resolution at five different bitrates using the Constant Bitrate (CBR) mode of encoding. Two different presets were used to simulate real world application requirements taking into account the codec complexity: medium preset for x264 and LCEVC-x264 and veryfast for x265 and LCEVC-x265. A summary of encoding settings is present in Table~\ref{tab:EncSettings}.
%
\begin{table}[t]
  \centering
  \caption{Encoding Settings Summary.}
    \resizebox{1.0\linewidth}{!}{
      \begin{tabular}{|l|l|}
    \toprule
    \textbf{Parameter } & \textbf{Value} \\
    \midrule
    Duration  & 30 sec \\
    \midrule
    Resolution  & 1080p \\
    \midrule
    Bitrates (kbps) & 800, 1500, 3000, 4000, 6000 \\
    \midrule
    Frame Rate  & 60 \\
    \midrule
    Encoder  & FFmpeg \\
    \midrule
    Encoding Mode  & CBR \\
    \midrule
    Video Compression Standards  & H.264, H.265, LCEVC H.264, LCEVC H.265 \\
    \midrule
    Preset  & medium and veryfast \\
    \bottomrule
    \end{tabular}%
  \label{tab:EncSettings}%

  }
\end{table}%

\section{Results} \label{sec:Results}

We present here the results in terms of Rate-Quality (distortion) curves as well as percentage \ac{BD-BR} savings. The quality was evaluated using two objective metrics: PSNR and VMAF. VMAF, proposed by Netflix in 2016, has shown to have a high correlation with subjective scores for a wide range of contents, including gaming content \cite{Barman2018NREvaluation}. The VMAF version used in this work is 
$1.5.2$ with the default model \textit{vmaf\_v0.6.1.pkl}.

\subsection{Rate-Quality Curves}

In Figure~\ref{fig:QualityBRcurves} we present the Quality (VMAF) vs. Bitrate plots for two of the sample video sequences - one of low content complexity (Bejeweled3) and one of high content complexity (BlackDesert) for both x264 and x265 codecs and their respective LCEVC implementations\footnote{Plots for additional games and other results are available in the open source dataset available at \cite{LCEVCDataset}}. It can be observed that for all four cases, LCEVC outperforms the respective base codec. 
While the quality gain for low complexity games is not that high, for high complexity games the visual quality gain is more evident, especially for the x264 case. The actual visual quality gain is quantified using  subjective ratings, discussed later in Section~\ref{subsec:Subjective}.

\subsection{BD-BR Analysis}

We calculated the results of percentage bitrate savings using the Python implementation available in \cite{pythonother} using the recommended piecewise fitting and support for 5 data points (DPs) \cite{MHV_BD_BR}. For brevity, we present here in Table~\ref{tab:BDBR-Results} only the bitrate savings using VMAF as 
quality metric. Other results, including BD-BR results using PSNR as the quality metric, are made available as part of the open source dataset \cite{LCEVCDataset}. 

From the results in Table~\ref{tab:BDBR-Results}, it is clear that, while using VMAF as the objective quality metric, LCEVC enhancing x264 (medium) and LCEVC enhancing x265 (veryfast) outperform both the base codecs, x264 and x265, used alone at full resolution, with BD-rate-VMAF of -42.14\% for x264 and -38.86\% for x265,  respectively. LCEVC-x264 (medium) outperforms x265 (veryfast) with a BD-BR-VMAF of -13.64\%. x265 results in 29.74\% higher
bitrate savings than x264 (considering VMAF). However, in terms of BD-BR analysis using PSNR, while LCEVC-x264 outperforms x264, x265 actually outperforms LCEVC-x265. Since in the end, the end user is the final judge of the actual visual quality, we performed subjective tests to obtain Mean Opinion Score (MOS) for different encoded video sequences to estimate the bitrate saving using MOS as the quality metric as well as compare the different encoded bitrate sequences for higher subjective visual quality.
\begin{table*}[t!]
  \centering
  \caption{BD-BR Bitrate Savings for different comparisons using VMAF as the Quality Metric.}
  \resizebox{0.75\linewidth}{!}{
      \begin{tabular}{|p{7.61em}|c|c|c|c|}
\cmidrule{2-5}    \multicolumn{1}{r|}{} & \multicolumn{4}{c|}{\textbf{BD-Rate-VMAF (\%)}} \\
    \midrule
    \multicolumn{1}{|l|}{\textbf{Video Sequence}} & \multicolumn{1}{p{6.835em}|}{\textbf{x264 vs LCEVC-x264}} & \multicolumn{1}{p{6.555em}|}{\textbf{x265 vs LCEVC-x265}} & \multicolumn{1}{p{7.445em}|}{\textbf{x265 vs LCEVC-x264}} & \multicolumn{1}{p{6.055em}|}{\textbf{x264 vs x265}} \\
    \midrule
    \multicolumn{1}{|l|}{Bejeweled3} & -69.02 & \cellcolor[rgb]{ 1,  .78,  .808}NaN & -70.92 & -2.50 \\
    \midrule
    \multicolumn{1}{|l|}{BlackDesert} & -42.15 & -39.63 & \cellcolor[rgb]{ 1,  .78,  .808}0.60 & -42.97 \\
    \midrule
    \multicolumn{1}{|l|}{Dauntless} & -40.15 & -35.73 & \cellcolor[rgb]{ 1,  .78,  .808}4.34 & -42.87 \\
    \midrule
    \multicolumn{1}{|l|}{ffxv} & -41.57 & -34.34 & -4.68 & -38.82 \\
    \midrule
    \multicolumn{1}{|l|}{Fortnite} & -41.56 & -35.80 & -2.51 & -40.25 \\
    \midrule
    \multicolumn{1}{|l|}{GTA5} & -43.02 & -41.52 & -6.60 & -39.30 \\
    \midrule
    \multicolumn{1}{|l|}{LoL} & -41.87 & -52.19 & -42.61 & -1.36 \\
    \midrule
    \multicolumn{1}{|l|}{maplestory} & -30.56 & -23.86 & \cellcolor[rgb]{ 1,  .78,  .808}0.20 & -29.81 \\
    \midrule
    \multicolumn{1}{|l|}{Minecraft} & -48.53 & -46.84 & -22.98 & -33.54 \\
    \midrule
    \multicolumn{1}{|l|}{Overwatch} & -43.77 & -38.07 & \cellcolor[rgb]{ 1,  .78,  .808}3.33 & -45.23 \\
    \midrule
    \multicolumn{1}{|l|}{R5Apex} & -36.74 & -36.72 & -4.54 & -33.83 \\
    \midrule
    \multicolumn{1}{|l|}{Rayman Legend} & -39.69 & -40.74 & -5.17 & -36.40 \\
    \midrule
    \multicolumn{1}{|l|}{Tekken} & -42.93 & -31.92 & -2.81 & -41.91 \\
    \midrule
    \multicolumn{1}{|l|}{Worms} & -28.35 & -47.81 & -36.66 & \cellcolor[rgb]{ 1,  .78,  .808}12.45 \\
    \midrule
    \rowcolor[rgb]{ 1,  1,  0} \textbf{Total BD-BR \newline Savings (\%)} & \textbf{-42.14} & \textbf{-38.86} & \textbf{-13.64} & \textbf{-29.74} \\
    \bottomrule
    \end{tabular}%
\label{tab:BDBR-Results}
 }
 \begin{tablenotes}
 \item Note: x264 and LCEVC-x264 are for videos encoded using medium preset, while x265 and LCEVC-x265 results are for videos encoded using veryfast preset. The coloured cells highlights the values different from general trend for the respective compared codecs. 
 \end{tablenotes}
\end{table*}

\subsection{Subjective Assessment} \label{subsec:Subjective}

In order to quantify the actual bitrate savings in terms of quality as perceived by the end user, a subjective test was conducted for multiple operating points for five source video sequences. Among fourteen video sequences, five video sequences, BlackDesert, Dauntless, GTA5, Overwatch, and Tekken, were selected because of their high spatial and temporal complexity. The high complex video games are selected since the low complexity games reach saturation in terms of visual quality at a low bitrate level. 

Due to Covid-19 restrictions, the subjective test was outsourced to an independent lab, VABTECH in Italy. The subjective test followed a Pairwise Comparison (PC) using a 7-point extended continuous (EC) Absolute Category Rating (ACR) rating scale adhering to the ITU-T Rec. P.809 subjective test guidelines \cite{ITU809} using a tool designed by TU Berlin. All the video sequences are available locally on the testing PC, which could then be played on a 24" PC Monitor using the Firefox browser. The subjective test consisted of four parts: a pre-test questionnaire, test instructions, rating job, and a post-test questionnaire. Before the test, the participants were instructed about the subjective test procedure and were allowed to ask any questions they might have. Since in gaming video quality tests participants can often mistake game design (graphic) quality for video quality, there was a training session with special instructions to make sure that participants do not judge the game design-related aspect but rather the video quality with visible compression artifacts i.e., blockiness and blur. This also helped the viewers  get familiar with the test procedure and rating scales. The training test consists of three pairs of videos displayed to subjects similar to the rating job, each pair with different quality levels and not from games used for the actual subjective tests. 

During the test, a pair of videos from the same codec family (e.g., x264 vs. LCEVC-x264 and x265 vs. LCEVC-x265) and same target bitrates were shown sequentially (in random order), and subjects were asked to rate both of them using two separate 7-point scales provided after watching both sequences. This allowed us to collect the subject's judgment about each video quality independently but also at the same time allow the participant to compare the two sequences. This allowed us to get insight into the improvement that LCEVC offers at different quality levels. A total of 16 test participants took part and there were no statistical outliers.

\vspace{-3mm}
\subsection{Subjective Assessment Results}

\begin{figure*}[h!]
    \centering
  \resizebox{0.7\linewidth}{!}{

    \begin{subfigure}[t]{0.5\textwidth}
    \includegraphics[ width=1.0\linewidth]{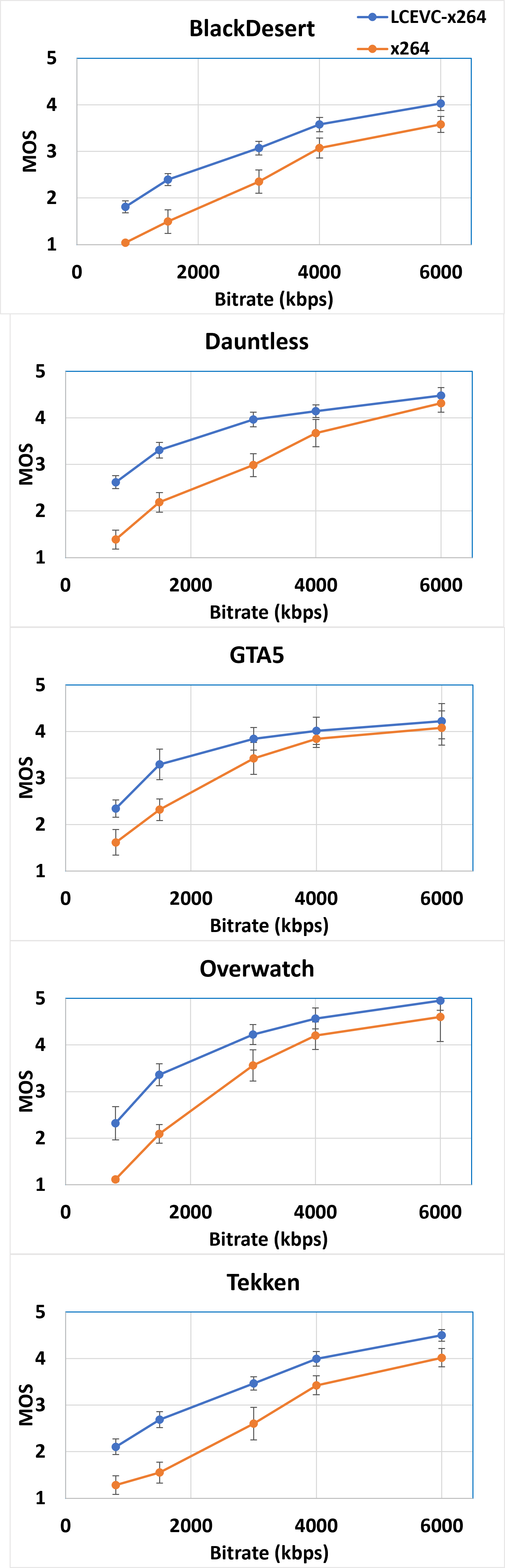}
    \caption{
    MOS with 95\% CI for x264 and LCEVC-x264.}
    \label{fig:bar_x264}
    \end{subfigure}%
    \begin{subfigure}[t]{0.5\textwidth}
    \centering
    \includegraphics[ width=1.0\linewidth]{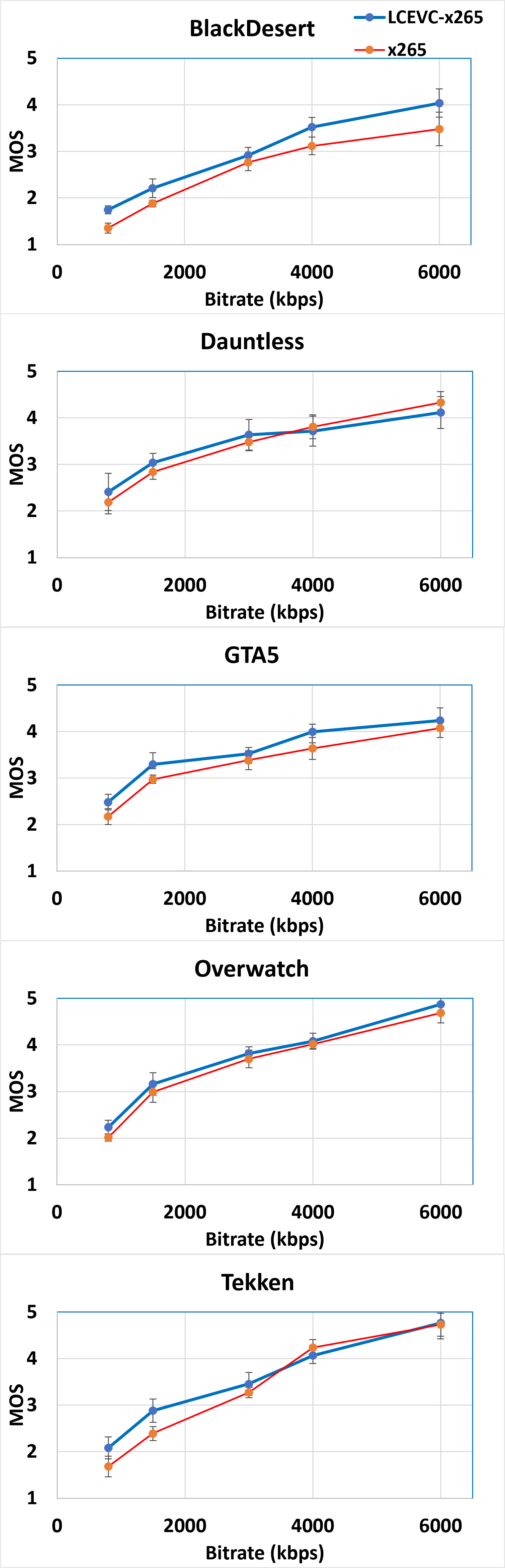}
    \caption{
    MOS with 95\% CI for x265 and LCEVC-x265.}
    \label{fig:bar_x265}
    \end{subfigure}
    }
    \caption{MOS vs. Bitrate (Kbps) plots for the five video sequences comparing performance of x264 . LCEVC-x264 and x265 vs. LCEVC-x265.}
    \label{fig:barplots}
\end{figure*}

Figure~\ref{fig:barplots} presents the MOS vs Bitrate plots for the five different games considering both compared codecs (x264 vs. LCEVC-x264 and x265 vs. LCEVC-x265) along with a 95\% confidence interval (CI) considering average MOS scores for all test subjects. It needs to be noted that the range of MOS scores is from 1-7 (extremely bad to excellent). However, none of the video sequences reached quality score higher than 5 (good). 

Looking at the plots, one can see that in general LCEVC results in  higher subjective scores compared to the base codec, especially at the lower bitrates. At higher bitrates, especially when considering x265 vs. LCEVC-x265, the absolute differences between the implementations are very small with the confidence interval often overalapping. For some sequences, such as Dauntless and Tekken, at 4 Mbps one can even see a reverse trend with x265 outperforming LCEVC-x265. 

To further understand and quantify the difference in performance of the compared codecs in terms of the effect of different factors on the observed results, a three-way analysis of variance (ANOVA) is conducted using the game, bitrate, and codec implementation (base codecs and their respective LCEVC implementations) as independent variables, and the video quality as the dependent variable. The analysis was performed for x264 and x265 separately. The test statistics of the ANOVA are summarized in Table \ref{tab:ANOVA_stats}.

\begin{table*}[!h]
\begin{center}
\caption{Test statistics of ANOVA for x264 vs LCEVC-x264 and x265 vs LCEVC-x265 using MOS scores.}
\label{tab:ANOVA_stats}
\resizebox{1.0\linewidth}{!}{
\begin{tabular}{@{}lllllllllll@{}}
\toprule
                               & \multicolumn{5}{c|}{x264}         & \multicolumn{5}{c}{x265}           \\ \midrule
Effect                         & F      & df1 & df2 & p     & \multicolumn{1}{c|}{$\eta_p^2$ }  & F       & df1 & df2 & p     & $\eta_p^2$  \\
\midrule
Game                           & 45.17  & 4   & 60  & \textless{}.001 & \multicolumn{1}{c|}{.75} & 48.26   & 4   & 60  & \textless{}.001 & .76 \\
Bitrate                        & 642.41 & 4   & 60  & \textless{}.001 & \multicolumn{1}{c|}{.98} & 1877.83 & 4   & 60  & \textless{}.001 & .99 \\
Codec Implementation           & 344.62 & 1   & 15  & \textless{}.001 & \multicolumn{1}{c|}{.96} & 31.59   & 1   & 15  & \textless{}.001 & .68 \\
Game x Bitrate                 & 9.84   & 16  & 240 & \textless{}.001 &\multicolumn{1}{c|}{.40} & 19.21   & 16  & 240 & \textless{}.001 & .56 \\
Game x Codec Implementation    & 2.91   & 4   & 60  & .029 & \multicolumn{1}{c|}{.16} & 4.93    & 4   & 60  & .002 & .25 \\
Bitrate x Codec Implementation & 33.40  & 4   & 60  & \textless{}.001 & \multicolumn{1}{c|}{.69} & 4.26    & 4   & 60  & .004 & .22 \\ \bottomrule
\end{tabular}}
\end{center}
\end{table*}

For x264, the ANOVA revealed a very strong major effect of the codec implementation and also an interaction effect of the codec implementation with the game as well as with the bitrate. As shown in Fig. \ref{fig:bar_x264}, LCEVC-x264 encoded videos are rated higher than the respective x264 encoded video sequences. However, this effect is much stronger for lower bitrates. For x265 and corresponding LCEVC-x265 encoded video sequence, the ANOVA also yielded a main effect of the codec implementation, but the effect size is much smaller. Once more, the differences between the codec implementations are higher for lower bitrates. It must also be reported that for one game, Dauntless, no statistically significant difference between the codecs was revealed. When considering the SI and TI values as depicted in Fig. \ref{fig:screenshots}, it turns out that this finding cannot be explained solely by taking into account the complexity of video sequences. 

For x265 vs. LCEVC-x265, Double Stimulus Impairment Scale (DSIS) subjective assessment was also performed by the same independent lab using ITU-T Rec BT.500-14 \cite{ITU-BT500-14} using a total of 20 test participant. The results (included in the dataset), indicate a better performance of LCEVC-x265 against x265, as observed using VMAF scores. For a more detailed analysis of the performance considering individual games and parameter levels, all RD plots along with subjective test scores are available in the dataset \cite{LCEVCDataset} and are not discussed here due to brevity.


\subsection{Performance of VQA}

LCEVC is an enhancement video coding codec where the aim is to provide a richer quality to the end users' perception
for constrained delay/complexity. While LCEVC is fundamentally different compared to super-resolution enhancement models, there are some shared similarities between their objectives, especially when considering the enhancement methodology. It has been argued in super-resolution works that full-reference metrics are not desired for quality measurement \cite{blau20182018}, as the image/video quality could get enhanced without necessarily becoming similar to the reference signal. This could also, on the other hand, lead to an inaccurate BD-BR analysis, which was presented earlier. Therefore, in this section, the performance of VMAF that is used for BD-BR analysis is evaluated together with PSNR, SSIM, MS-SSIM and a no-reference gaming quality metric, NDNetGaming \cite{utke2020ndnetgaming}. Table \ref{tab:PCC} suggests a high 
correlation of VMAF 
with subjective results
for all stimuli, with Pearson Linear Coefficient Correlation (PLCC) of $0.891$. However, the PLCC score is higher for native codecs as compared to the respective LCEVC implementation. Generally, PSNR and SSIM performs poor in terms of correlation with subjective scores, MS-SSIM performs reasonably, and the performance of VMAF and NDNetGaming depends on the codecs. Interestingly, NDNetGaming has a higher PLCC score for LCEVC \acp{PVS} compared to VMAF for x265 \acp{PVS} (considering both native and LCEVC). VMAF performs better for x264 \acp{PVS} (both LCEVC and x264 combined) and the native codecs. While the difference compared to VMAF is small, it should be noted that NDNetGaming is a no-reference metric as compared to VMAF which is a full-reference metric.

\begin{table*}[]
 \centering
    \caption{PLCC for three quality metrics considering different sets of PVSs of the subjective test dataset.}
  \resizebox{0.7\linewidth}{!}{
\begin{tabular}{|c|ccccc|}
 \hline
\textbf{PVSs}                                                              & \multicolumn{1}{c|}{\textbf{PSNR}} & \multicolumn{1}{c|}{\textbf{VMAF}} & \multicolumn{1}{c|}{\textbf{SSIM}} & \multicolumn{1}{c|}{\textbf{MS-SSIM}} & \textbf{NDNetGaming}                 \\ \hline
\textbf{\begin{tabular}[c]{@{}c@{}}LCEVC-x264\\ LCEVC-x265\end{tabular}}   & \multicolumn{1}{c|}{0.59}          & \multicolumn{1}{c|}{0.85}          & \multicolumn{1}{c|}{0.63}          & \multicolumn{1}{c|}{0.79}             & 0.90                 \\ \hline
\textbf{\begin{tabular}[c]{@{}c@{}}Native-x264\\ Native-x265\end{tabular}} & \multicolumn{1}{c|}{0.73}          & \multicolumn{1}{c|}{0.91}          & \multicolumn{1}{c|}{0.68}          & \multicolumn{1}{c|}{0.83}             & 0.88                 \\ \hline
\textbf{\begin{tabular}[c]{@{}c@{}}LCEVC-x264\\ Native-x264\end{tabular}}  & \multicolumn{1}{c|}{0.73}          & \multicolumn{1}{c|}{0.94}          & \multicolumn{1}{c|}{0.69}          & \multicolumn{1}{c|}{0.86}             & 0.91                 \\ \hline
\textbf{\begin{tabular}[c]{@{}c@{}}LCEVC-x265\\ Native-x265\end{tabular}}  & \multicolumn{1}{c|}{0.56}          & \multicolumn{1}{c|}{0.86}          & \multicolumn{1}{c|}{0.61}          & \multicolumn{1}{c|}{0.78}             & 0.93                 \\ \hline
\textbf{All}                                                               & \multicolumn{1}{c|}{\textbf{0.65}}          & \multicolumn{1}{c|}{\textbf{0.89}}          & \multicolumn{1}{c|}{\textbf{0.66}}          & \multicolumn{1}{c|}{\textbf{0.82}}             & \textbf{0.88} 
\\ \hline
\end{tabular}}
\label{tab:PCC}
\end{table*}

\begin{figure*}[t!]
\begin{center}
\includegraphics[ width=0.9\linewidth]{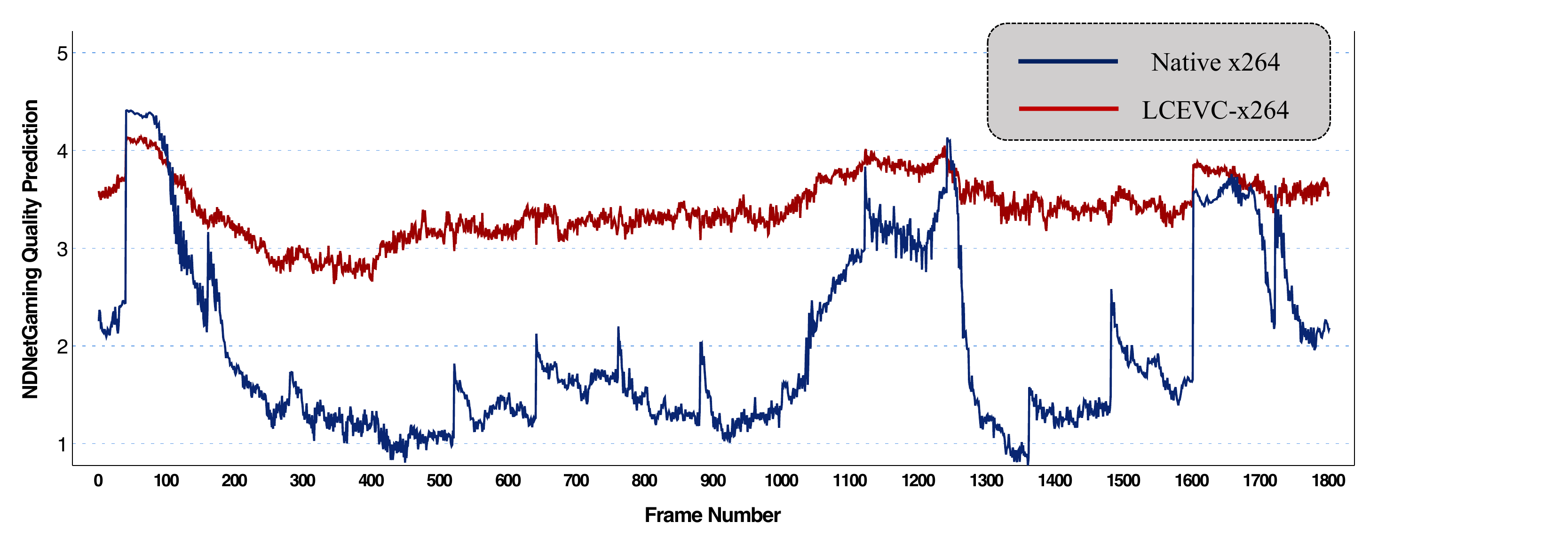}
\end{center} 
\caption{Per-frame quality of NDNetGaming scores for the entire GTA5 sequence encoded at 800 kbps.}
\label{fig:temporal}
\end{figure*}

\subsection{Temporal Quality Variation}

Gaming content is a special type of content that could include sudden high intensity motion instances, which makes it difficult for conventional video coding's rate controller to balance out the quality over a certain period. As an example of such quality changes, as it can be seen in Figure \ref{fig:temporal}, frame-level quality changes could be significant for games with high temporal complexity, as shown in the case of GTA gameplay. Figure \ref{fig:temporal} shows the temporal changes over time for a video sequence of GTA encoded at 800 kbps using H.264 medium preset. As it can be seen, due to sudden temporal changes, the quality varies significantly. Such a behavior can be seen less strongly for the HEVC codec or when the target bitrate is above a certain threshold. In this work, we observed that LCEVC also benefits gaming streaming by providing a balance of quality distribution, as can be seen in Figure \ref{fig:temporal}. Using NDNetGaming prediction (at a five-point MOS scale), the standard deviation of frame-level quality dropped from 0.41 when LCEVC is not used, to 0.23, when LCEVC is used. If only 25\% of the high temporal variation is taken into consideration, the gain increases from the standard deviation of $0.57$ to $0.28$ using NDNetGaming scores. Similar observation was obtained when analyzing the per-frame VMAF values. Lower quality variation is typically in favor of service providers as it can allow them providing almost uniform quality to their users. 

\subsection{Temporal Bitrate Variation}

\begin{figure*}[t!]
    \centering
    \begin{subfigure}[t]{0.92\textwidth}
    \includegraphics[ width=1.0\linewidth]{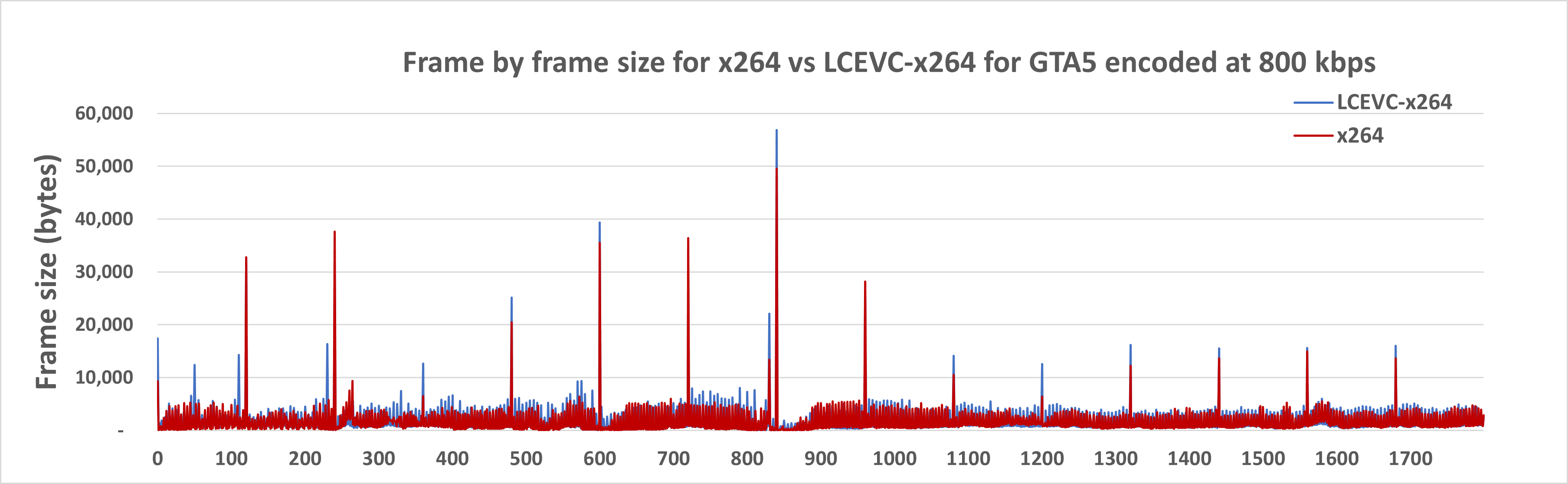}
    \label{fig:bar_x264}
    \end{subfigure}%
    
    \begin{subfigure}[t]{0.92\textwidth}
    \centering
    \includegraphics[ width=1.0\linewidth]{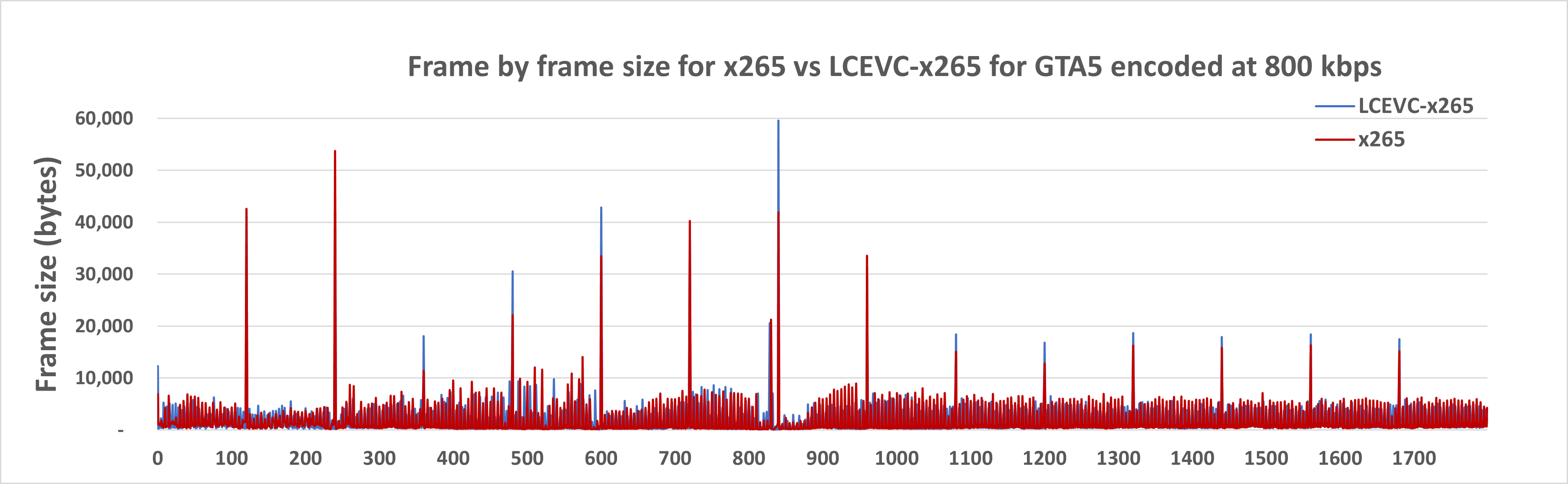}
    \end{subfigure}
    \caption{Frame by frame size for for GTA5 encoded at 800 kbps considering x264 vs LCEVC-x264 and x265 vs LCEVC-x265.}
    \label{fig:fbf}
\end{figure*}

While providing uniform quality is of interest to many service providers, they might be more interested in providing uniform rates to ensure smooth stream delivery.  
Therefore, in this section
we extracted the frame-by-frame size for the tested sequences using FFprobe. This would allow us to analyse whether such improvement in quality variation would result in bitrate variation or not. 
Figure~\ref{fig:fbf} shows the frame-by-frame plot for GTA5 800 Kbps for both x264/LCEVC-x264 and x265/LCEVC x265. It can be observed that the frame sizes are substantially aligned with no major differences for LCEVC encoded sequences (LCEVC-x264 and LCEVC-x265) as compared to the respective base codec (x264 and x265) encoded sequences. LCEVC's more consistent behaviour is due to the multi-layer nature of the codec: the base layer, being a half-resolution, can afford to encode at lower \ac{QP} vs. the same codec used at full resolution, as such it maintains a more balanced quality of experience even in  scarce bitrate situations.

Similar figures considering other games and bitrates along with frame-by-frame size for different encoded video sequences are made available in the dataset \cite{LCEVCDataset}. 

\subsection{Impact of Subjective Test Design}

In this work, a unique subjective test design is followed, which allowed us to benefit from both the ACR scale design as well as the pairwise comparison test design. However, this requires more cognitive efforts from the users to rate the stimuli independently while reflecting their pairwise comparison between the double stimuli of the native codec and their respective LCEVC implementation encoded sequence. Our post-test interview with test participants indicated that this led to a stricter judgment. Also, while analyzing the validity of such measurement requires an in-depth analysis and comparison to other methods, the correlation results with the different metrics suggest a very similar performance of VMAF and NDNetGaming as was observed in the original dataset, CGVDS \cite{SamanCGVDS}, from which we used the test sequences. Therefore, we can conclude that the subjective test method does not introduce significant errors to the results. However, the chosen video sequences due to their high temporal complexity might result in temporal masking and hence the results might vary depending on the choice of video sequences. In order to supplement the current results, an additional subjective test for x265 vs. LCEVC-x265 video sequences was carried out using the ITU BT.500 DCR test methodology using 20 test participants. The MOS scores (0-10 scale) are made available in the dataset \cite{LCEVCDataset}.  



\section{Conclusion and Future Work} \label{sec:Conc}

In this paper, we presented an extensive evaluation of the newly proposed MPEG-5 Part 2 LCEVC codec for live gaming video streaming applications. Two very widely used base codecs, x264 and x265, and their respective LCEVC implementations were compared both objectively and subjectively. We observed that when using VMAF as the quality metric, LCEVC outperformed the respective base codecs in terms of BD-BR results. However, in terms of PSNR, while LCEVC-x264 outperformed x264, x265 outperformed LCEVC-x265. In order to measure actual visual gains at the end-user, a paired comparison subjective test was performed which indicated that while LCEVC-264 indeed outperforms x264 in terms of overall MOS scores for the bitrates considered, the gain is more significant at lower bitrates. When comparing x265 to LCEVC-x265, we found that while RD curves indicated not so significant performance gains for LCEVC-x265, especially at high bitrates, the ANOVA analysis does indicate a main effect of the codec implementation. Given that the original BD-BR metric was proposed to be used with PSNR as the objective metric and as indicated by the results in this work, BD-BR results using other metrics such as VMAF should be interpreted with caution and, when possible, supplemented by subjective tests results. It should be noted that the results presented in this work are considering cloud gaming streaming applications, and the videos and settings used for encoding videos are representative of the same. Also, for the videos considered for the subjective tests, we observe that none of the videos was scored higher than 5 on a quality scale from 1-7. This, along with consideration of the fact that the chosen video sequences are of high frame rate and quite complex, further studies with higher bitrate ranges will give more realistic savings figures for the compared codes. As future work, we plan to extend this work to other codecs such as VP9, AV1, and VVC as based codecs. Our additional future work will include more studies considering multiple resolution-bitrate pairs and quantifying the performance on 4K, HDR games using the recently designed GamingHDRVideoSET \cite{GamingHDRVideoSET}.

\section*{Acknowledgements}
The authors would like to thank the team at V-Nova, especially Guendalina Cobianchi, Florian Maurer, Lorenzo Ciccarelli, Harry Morgan, Lorenzo Cassina and Simone Ferrara for providing us with the LCEVC SDK releases and help with setting up of the LCEVC encoding commands. Authors would also like to thank Vittorio Baroncini from VABTECH for conducting the subjective tests.

\bibliographystyle{IEEEtran}
\bibliography{ms.bib}
\begin{IEEEbiography}[{\includegraphics[width=1in,height=1.25in,clip,keepaspectratio]{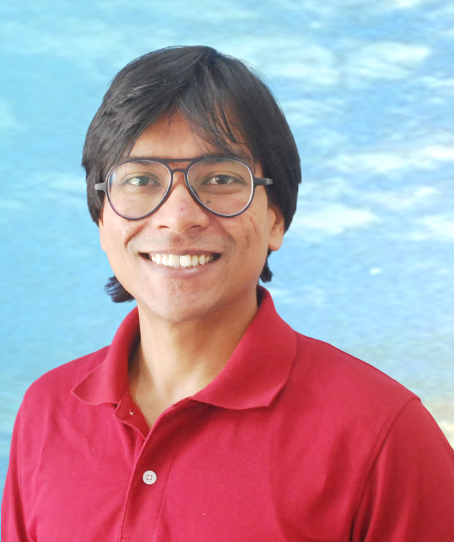}}]{Nabajeet Barman} (M'19) is currently a Principal Video Systems Engineer, Research at Brightcove where he works on optimal video encoding strategies, perceptual quality assessment and end-to-end optimization. He is also a Fellow of the Higher Education Academy (FHEA), UK and holds an adjunct Lecturer in Applied Computer Science position at Kingston University. He received his PhD and MBA from Kingston University, London and MSc in IT from Universität Stuttgart, Germany and B.Tech in Electronics Engineering from NIT, Surat, India. Previously, he was a full-time Lecturer in Applied Computer Science (Data Science) with Kingston University, London. From 2012-2015, he worked in different capacities across various industries including Bell Labs, Stuttgart, Germany after which he joined Kingston University as a Marie Curie Fellow with MSCA ITN QoE-Net from 2015 to 2018, and a Post-Doctoral Research Fellow from 2019-2020. He is a Board Member of the Video Quality Expert Group (VQEG) as part of the Computer Graphics Imaging (CGI) group and is a member of various standardization groups such as ITU-T SG16 and MPEG, SVA, CTA, DASH-IF and UHDF. He has published in many international conferences and journals and is an active reviewer for many conferences and journals. 
\end{IEEEbiography}

\begin{IEEEbiography}[{\includegraphics[width=1in,height=1.25in,clip,keepaspectratio]{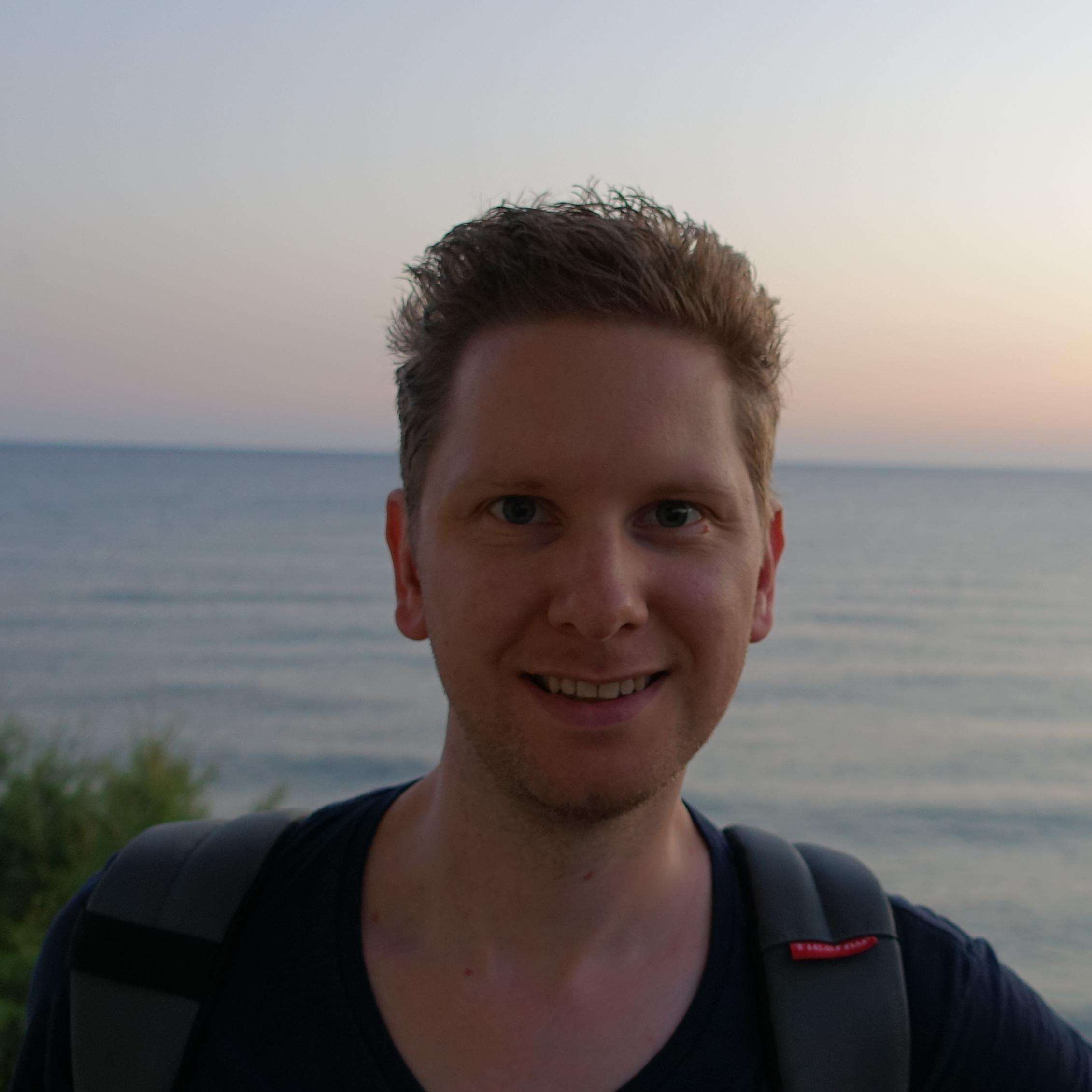}}]{Steven Schmidt}
Steven Schmidt received his Master degree in Electrical Engineering at the Technische Universität Berlin with a major in Communication Systems. From 2016 to 2021, he was employed as a research assistant at the Quality and Usability Lab where he completed his PhD in the field of Quality of Experience of cloud gaming services. Since 2022, he continued working in this domain at Sony PlayStation as a researcher and data scientist. His research interests include assessment methods for gaming QoE, the identification of influencing factors as well as developing models to predict gaming QoE of cloud gaming services. Therefore, he is involved in gaming-related activities of the ITU Telecommunication Standardization Sector (ITU-T).
\end{IEEEbiography}

\begin{IEEEbiography}[{\includegraphics[width=1in,height=1.25in,clip,keepaspectratio]{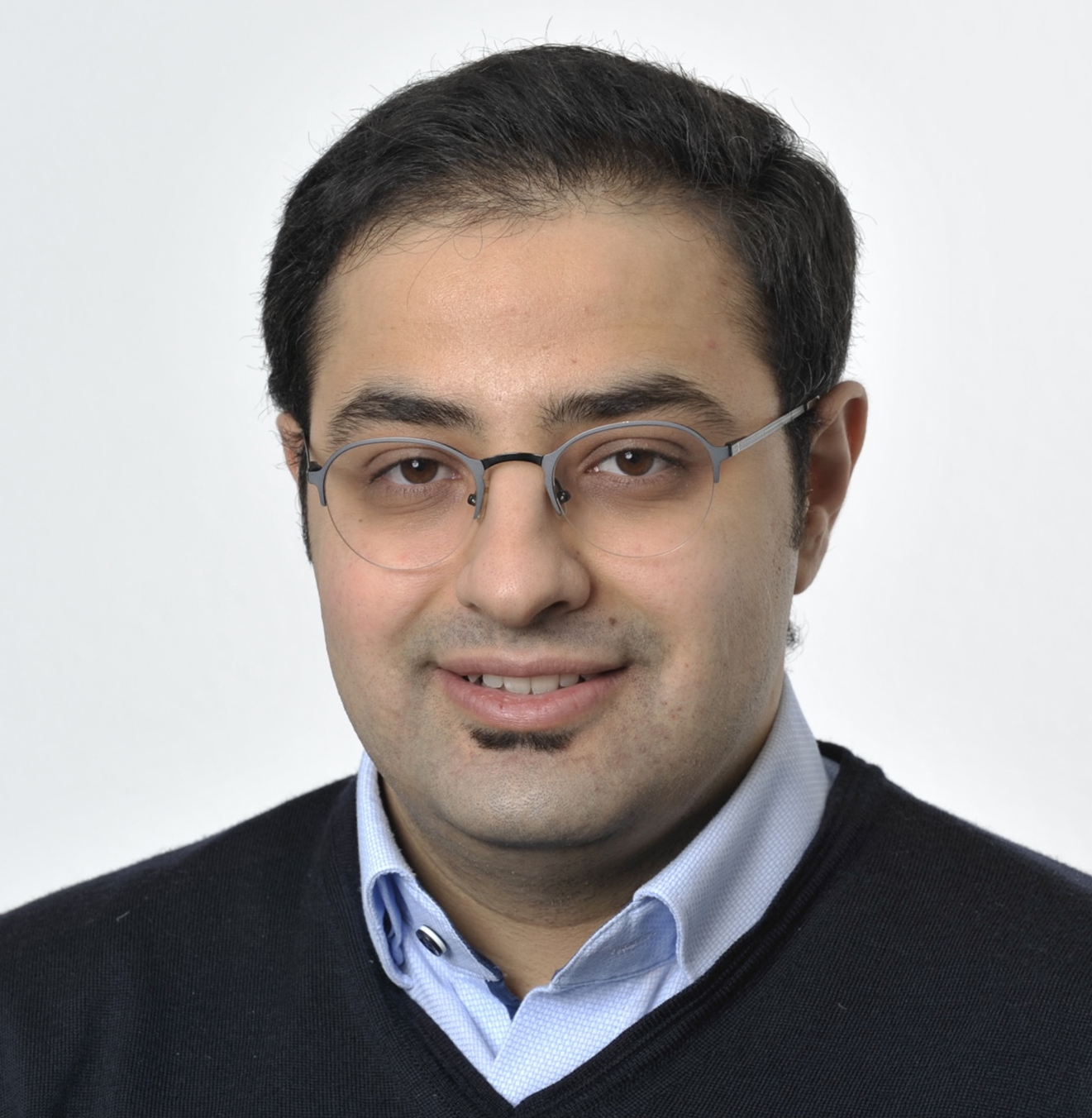}}]{Saman Zadtootaghaj} is currently a senior research engineer at Dolby Laboratories. His main interest is the subjective and objective quality assessment of computer-generated content. He worked as a researcher at Telekom Innovation Laboratories of Deutsche Telekom AG from 2016 to 2018 as part of the European project called QoE-Net. He completed his Ph.D. at the Quality Usability lab group of TU Berlin in 2021. He is currently the chair of the Computer-Generated Imagery group at Video Quality Expert Group and an active member of ITU-T Study Group 12, leading the P.BBQCG work item. He strongly contributed to the ITU-T SG 12, including over 20 contributions within the past five years.
\end{IEEEbiography}

\begin{IEEEbiography}[{\includegraphics[width=1in,height=1.25in,clip,keepaspectratio]{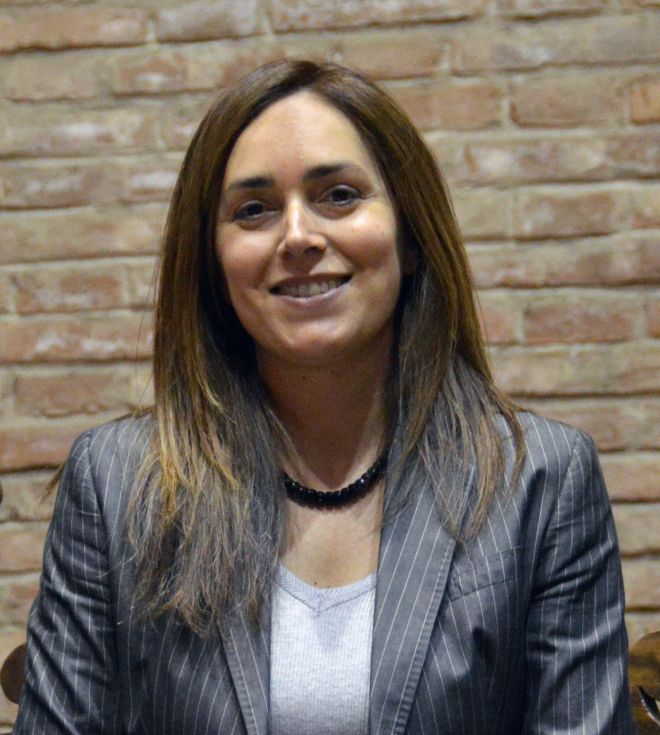}}]{Maria Martini}
 [SrM’07] is Professor in the Faculty of Science, Engineering and Computing at Kingston University, London, U.K., where she also leads the Wireless Multimedia Networking Research Group and she is the Course Director for the MSc in “Networking and Data Communications.” She is a Fellow of The Higher Education Academy (HEA). She received the Laurea degree in electronic engineering (summa cum laude) from the University of Perugia, Italy, in 1998 and the PhD degree in Electronics and Computer Science from the University of Bologna, Italy, in 2002. She has led the KU team in a number of national and international research projects, funded by the European Commission (e.g., OPTIMIX, CONCERTO, QoE-NET, Qualinet), U.K. research councils (e.g., EPSRC, British Council, Royal Society), Innovate UK, and international industries.  Associate Editor for IEEE Signal Processing Magazine (2018-2021) and  IEEE Transactions on Multimedia (2014-2018), she was lead guest editor for the IEEE JSAC special issue on "QoE-aware wireless multimedia systems" (2012), and  editor for IEEE Journal of Biomedical and Health Informatics (2014), IEEE Multimedia (2018),  Int. Journal on Telemedicine and Applications, among others. Expert Evaluator and Panel Member for the European Commission and for national funding agencies (e.g. EPSRC in the UK), she  is part of the NetWorld Europe ETP Expert Group, Board member of the Video Quality Expert Group (VQEG) and member of the IEEE Multimedia Communications technical committee, currently serving in the Awards Committee and having served as vice-chair (2014-2016), as chair  of the 3D Rendering, Processing, and Communications Interest Group (2012-2014), and as key member of the QoE and multimedia streaming IG. Her research interests include wireless multimedia networks, video quality assessment, decision theory, machine learning, and medical applications. She authored about 200 international scientific articles and  book sections,  international patents and contributions to international standards (IEEE and ITU). She currently chairs the IEEE P3333.1.4 standardization working group on the quality assessment of light field imaging. 
\end{IEEEbiography}
\end{document}